\documentclass{article}

\usepackage{geometry}
\usepackage{newtxtext,newtxmath}

\usepackage[numbers,sort&compress]{natbib}
\usepackage{microtype}
\usepackage{graphicx}
\usepackage{subcaption}
\usepackage{booktabs} 

\usepackage{amsmath}
\usepackage{mathtools}
\usepackage{authblk}
\usepackage{threeparttable}
\usepackage{xcolor}
\usepackage{colortbl}
\usepackage{pifont}
\usepackage{multirow}
\usepackage{float}
\usepackage[most]{tcolorbox}
\usepackage{algorithm}
\usepackage{algorithmic}

\usepackage[most]{tcolorbox}
\newtcolorbox{promptbox}{
  colback=orange!10,
  colframe=orange!50,
  arc=6pt,
  boxrule=0.8pt,
  left=6pt,
  right=6pt,
  top=6pt,
  bottom=6pt
}

\title{ProphetKV: User-Query-Driven Selective Recomputation for Efficient KV Cache Reuse in Retrieval-Augmented Generation}

\author[1]{Shihao Wang\thanks{These authors contributed equally to this work.}}
\author[1]{Jiahao Chen\protect\footnotemark[1]}
\author[1]{Yanqi Pan}
\author[1]{Hao Huang}
\author[1]{Yichen Hao}
\author[1]{Xiangyu Zou\thanks{Corresponding authors: \{zouxiangyu, wenxia\}@hit.edu.cn}}
\author[1]{Wen Xia\protect\footnotemark[2]}
\author[2]{Wentao Zhang}
\author[2]{Chongyang Qiu}
\author[2]{Pengfei Wang}

\affil[1]{Harbin Institute of Technology, Shenzhen}
\affil[2]{Beijing Yanrong Technology Co., Ltd.}

\begin{document}
\date{}
\maketitle

\begin{abstract}
The prefill stage of long-context Retrieval-Augmented Generation (RAG) is severely bottlenecked by computational overhead. To mitigate this, recent methods assemble pre-calculated KV caches of retrieved RAG documents (by a \emph{user query}) and reprocess selected tokens to recover cross-attention between these pre-calculated KV caches. However, we identify a fundamental ``crowding-out effect'' in current token selection criteria: globally salient but \emph{user-query}-irrelevant tokens saturate the limited recomputation budget, displacing the tokens truly essential for answering the \emph{user query} and degrading inference accuracy.

We propose ProphetKV, a user-query-driven KV Cache reuse method for RAG scenarios. ProphetKV dynamically prioritizes tokens based on their semantic relevance to the \emph{user query} and employs a dual-stage recomputation pipeline to fuse layer-wise attention metrics into a high-utility set. 
By ensuring the recomputation budget is dedicated to bridging the informational gap between retrieved context and the \emph{user query}, ProphetKV achieves high-fidelity attention recovery with minimal overhead. Our extensive evaluation results show that ProphetKV retains 96\%–101\% of full-prefill accuracy with only a 20\% recomputation ratio, while achieving accuracy improvements of 8.8\%–24.9\% on RULER and 18.6\%–50.9\% on LongBench over the state-of-the-art approaches (e.g., CacheBlend, EPIC, and KVShare).
\end{abstract}

\section{Introduction}

Large Language Models (LLMs) integrated with Retrieval-Augmented Generation (RAG) have become the \emph{de facto} standard for addressing domain-specific tasks~\cite{rag-1, rag-2, rag-3}. 
In a typical RAG pipeline, the system retrieves a collection of relevant document chunks from a vast external corpus based on a \emph{user query}, which are then concatenated to form the input context. However, to ensure comprehensive grounding and high answer quality, modern RAG systems often need to process an increasing number of retrieved fragments, scaling the total context length from 10k to 1M tokens~\cite{rag-lc-1, rag-lc-2, cacheblend}.

\begin{figure}[t]
    \begin{center}
      \centerline{\includegraphics[width=\columnwidth]{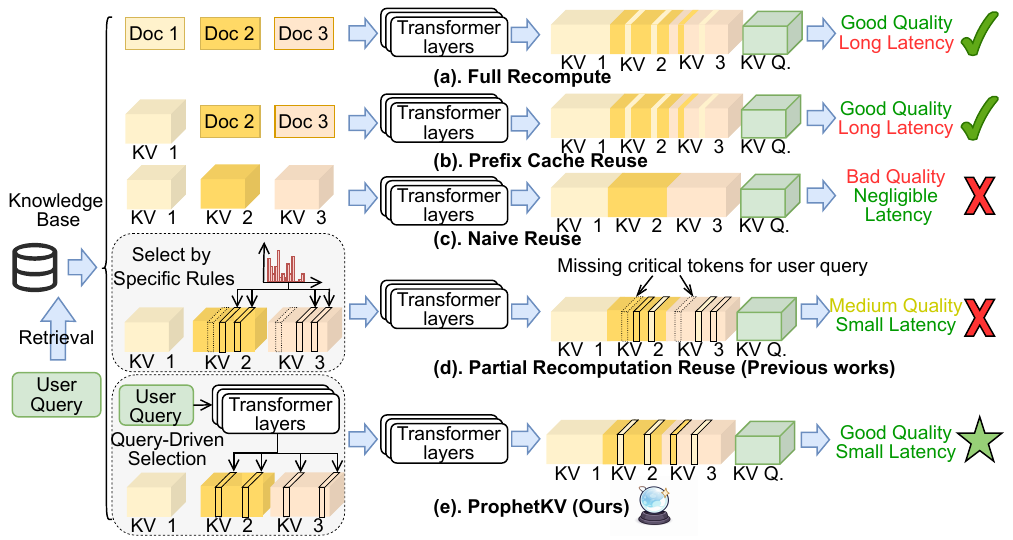}}
      \vspace{-4pt}
      \caption{
        A high-level overview of different methods of KV Cache reuse in RAG scenarios.
      }
      \label{fig: intro}
    \end{center}
    \vspace{-30pt}
\end{figure}

This massive input size imposes a severe computational burden, particularly during the prefill stage of LLM inference. While LLM inference is generally divided into a prefill stage (processing the prompt) and a decode stage (generating tokens), the former becomes the primary bottleneck in long-context RAG scenarios. Unlike the autoregressive decode stage, the prefill stage must compute self-attention across the entire sequence to aggregate semantic information into a Key-Value (KV) cache. As the computational complexity of the attention mechanism scales as $O(N^2)$ with sequence length $N$, the prefill latency (i.e., Time-to-First-Token, TTFT) becomes prohibitively high, severely hindering the responsiveness of real-time RAG services.

To alleviate the prefill bottleneck, KV cache reuse has emerged as a cornerstone strategy. Traditional methods rely on strict prefix matching~\cite{sglang}, which constrains reuse to identical sequences. However, this requirement is rarely satisfied in RAG scenarios, where retrieved documents are dynamically reordered and seldom share a common prefix~\cite{cacheblend}. Consequently, recent research has shifted toward position-independent (PI)\footnote{To ensure position independence, positional encodings are excluded from precomputed Key caches and restored at runtime \cite{cacheblend}. We thus treat this mechanism as given and focus on the reuse strategy.} KV cache reuse~\cite{cacheblend, epic}, enabling the assembly of precomputed chunk-wise caches regardless of their original sequence order.

However, simply concatenating KV caches that were precomputed in isolation causes severe accuracy degradation. This is because such an approach neglects \textbf{cross-attention} between documents: they have never ``seen" each other during precalculation. To reinstate these inter-document dependencies, state-of-the-art methods~\cite{cacheblend, epic} introduce partial recomputation, selectively recomputing the KV cache for a small subset of tokens to ``bridge'' the resulting semantic fragmentation and strike a balance between calculation saving and inference accuracy.

Despite these efforts, we identify a fundamental flaw in existing recomputation schemes: their selection criteria are inherently blind. By relying on global attention weights or positional heuristics, these methods strive to approximate the full cross-attention map of a standard Transformer prefill. However, they fail to distinguish between generic saliency and task-specific relevance: only a sparse subset of pivotal sentences is task-critical in retrieved documents, whereas the vast other content remains semantically extraneous to the specific \emph{user query}. Reconstruction cross-attention for these ``unused'' tokens provides marginal utility for final answer generation but incurs a substantial recompute cost.
This induces a ``crowding-out effect'': the limited recomputation budget is saturated by globally active but task-irrelevant tokens, thereby displacing the really essential ones for accurate answer generation. As a result, we observe that existing methods suffer accuracy drops of up to 86\% on representative benchmarks (See Sec.~\ref{sec:evaluation-accuracy}), which limits their viability for real-world deployments and motivates the design of higher-fidelity recomputation methods.

Accordingly, we encapsulate our findings into \textbf{two core insights}: 
(1) The reconstruction objective adopted by prior methods is functionally redundant for RAG tasks; (2) The utility of cross-attention is strictly query-contingent. In RAG scenarios, the user query serves as a decisive semantic prior that defines the tokens' relevance.

Based on these insights, we propose ProphetKV, a user query-driven selective recomputation framework (Fig.~\ref{fig: intro}). ProphetKV facilitates a paradigm shift from blind approximation of the global attention landscape to query-targeted attention recovery, ensuring that partial recomputation mechanisms are dedicated exclusively to repairing cross-attention relevant to the specific query. 
ProphetKV optimizes the accuracy–efficiency trade-off through two synergistic components: (1) User Query-Guided Token Selection: We introduce a mechanism that leverages the attention weights of the user query to dynamically isolate context tokens that are semantically pivotal for the user query.
(2) Dual-Stage Recomputation with Layer Fusing: To address the challenge of inter-layer attention variance (where critical tokens shift in different Transformer layers), we propose a fusion algorithm. This fuser aggregates attention-related metrics across all layers to derive a unified, high-utility recomputation set that ensures consistent accuracy recovery.

Experimental results in Sec.~\ref{sec:evaluation-accuracy} demonstrate that, under a constrained budget of 20\% recomputed tokens, our method achieves significant accuracy gains of 8.8-24.9\% on the RULER dataset and 18.6-50.9\% on the LongBench dataset over prior SOTA approaches. Notably, our approach is the only strategy capable of nearing full recomputation accuracy, underscoring its potential to bridge the gap between theoretical efficiency and production-grade reliability.

\section{Background}
\label{sec:background}
\subsection{KV Cache in the Transformer architecture}
\label{sec:background-kv-cache}

Transformers aggregate contextual information via causal self-attention, where each token is associated with Query (Q), Key (K), and Value (V) representations. During inference, the model initiates a prefill stage to process the input prompt, caching the resulting Key and Value tensors (KV cache), which are reused to facilitate efficient token generation in the subsequent decoding stage. The model is organized as a stack of layers, where the output of each layer serves as the input to the next, forming a hierarchical dependency that captures long-range and abstract semantics.

\subsection{KV Cache Reuse and RAG applications}
\label{sec:background-kv-cache-reuse}
RAG augments LLMs with externally retrieved document chunks and prepends them to the input prompt to guide generation, resulting in long input sequences. Since retrieved chunks are frequently identical across requests, reusing their KV caches offers a promising opportunity to reduce prefill latency. Existing inference systems such as vLLM~\cite{vllm} and SGLang~\cite{sglang} employ prefix-based KV cache reuse, which reuses KV caches only when two requests share an identical prompt prefix. However, this strict requirement is ill-suited for RAG: even if two requests share identical chunks, any variation in their ordering breaks the prefix chain, rendering the KV cache non-reusable.

\begin{figure*}[t]
    \begin{center}  
        \centerline{\includegraphics[width=\textwidth]{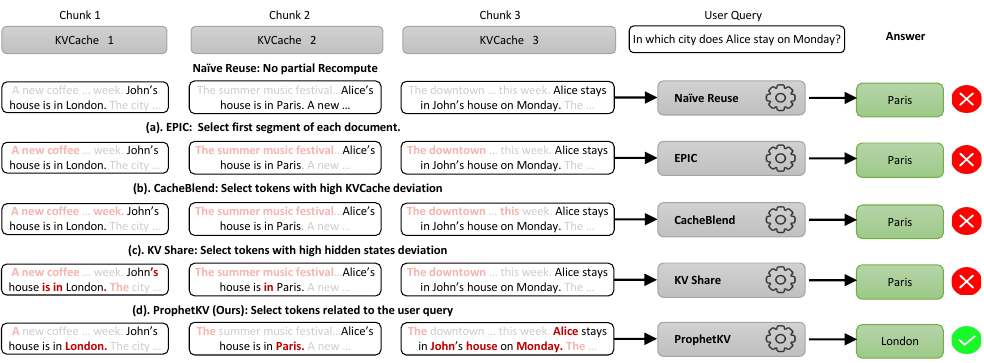}} 
        \vspace{-7pt}
        \caption{An example illustrating the selected tokens of existing approaches under a 20\% recomputation ratio. Text irrelevant to the user query is colored in gray, and tokens selected by each method are colored in red. See Appendix~\ref{sec:appendix-example-prompt} for the full prompt.} 
        \label{fig:Moti-Example} 
    \end{center} 
    \vspace{-30pt}
\end{figure*}

To address this limitation, position-independent (PI) KV cache reuse decouples precomputed KV caches from their absolute token positions. A naïve PI reuse strategy directly concatenates precomputed chunk caches, which maximizes computational savings but degrades accuracy due to missing cross-attention interactions. Recent work alleviates this issue via partial recomputation, which selectively recomputes a small subset of tokens to reconstruct cross-attention. 

Existing methods fall into two categories: \textbf{training-free} and \textbf{fine-tuned} approaches. Training-free methods are classified as static or dynamic, depending on whether token selection depends on the input prompt. EPIC~\cite{epic} is a typical static method based on the attention-sink phenomenon to select the initial tokens of each chunk for recomputation. Methods based on dynamic rules include CacheBlend~\cite{cacheblend} and KVShare~\cite{kvshare}; these derive token-selection rules from numerical error analyses of deviations in the KV cache and in hidden states, respectively. Fine-tuned methods are represented by CacheClip~\cite{cacheclip}, which fine-tunes a small auxiliary model to predict recomputation-worthy tokens by exploiting similarity between the auxiliary and target models. While effective under controlled settings, such approaches suffer from heavy run-time overhead and limited practicality in open-domain or rapidly evolving workloads. 

Given the practical limitations of fine-tuned approaches, this work focuses on training-free, plug-and-play partial recomputation for KV cache reuse. We do not consider finetuned solutions, as our goal is to preserve the native generalization of LLMs and avoid additional training. Nevertheless, existing training-free partial recomputation methods still struggle to achieve a satisfactory balance between accuracy and computational efficiency on various benchmarks (see Sec.~\ref{sec:evaluation}), indicating substantial room for further improvement.
\section{Motivation} 
\label{sec:motivation}
\subsection{Illustrating the Failure of Existing Methods} 
\label{sec:motivation-failure}

Existing partial-recomputation methods aim to reconstruct the entire missing cross-chunk attention under the assumption that this is feasible within a strict budget. However, our evaluation in Sec.~\ref{sec:evaluation-accuracy} reveals a consistent failure to maintain accuracy in various RAG scenarios.

\begin{figure}[t]
    \begin{center} 
        \centerline{\includegraphics[width=0.5\columnwidth]{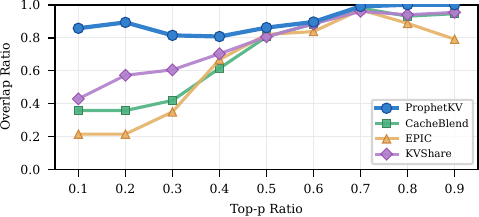}} 
        \vspace{-5pt}
        \caption{Overlap ratio ($|S \cap G| / |G|$) between selected tokens ($S$) and query-attended tokens ($G$), where $G$ is the average query-to-context attention under full-prefill, across selection ratios ($p$).} 
        \label{fig:Moti-topp-overlap} 
    \end{center}
    \vspace{-36pt}
\end{figure}

We analyze a representative RAG case in which a query, ``In which city does Alice stay on Monday?'' requires bridging information between Chunks 1 and 3, whereas Chunk 2 contains misleading details. As shown in Fig.~\ref{fig:Moti-Example}, all existing methods fail to reconstruct this significant cross-attention and thus predict an incorrect answer, i.e. ``Paris''. Notably, the selected tokens of these methods do not align with user-query-relevant tokens identified by human intuition. We quantify this mismatch by measuring the overlap ratio, an indicator widely used in prior works~\cite{pyramidinfer, cacheclip}, to evaluate the consistency between selected tokens and query-critical tokens in Fig.~\ref{fig:Moti-topp-overlap}. The combined results of Fig.~\ref{fig:Moti-Example} and Fig.~\ref{fig:Moti-topp-overlap} suggest that \textbf{the impact of missed user query-related tokens leads to incorrect predictions}.

\textbf{Degradation of Current Recomputation Methods.} EPIC relies on static heuristics, which lack the dynamic, input-dependent flexibility of Transformers. CacheBlend and KVShare utilize deviation-based criteria that, while theoretically motivated, are difficult to estimate without a full prefill pass. To reduce overhead, they approximate high-layer dependencies using low-layer information. However, since low-layer representations are often misaligned with high-layer semantics, these methods fail to capture tokens whose relevance only emerges in deeper layers(See Sec.~\ref{sec:design-recomputation}). As a natural result, these methods fail to recover the whole missing cross-attention, resulting in the failure in Fig.~\ref{fig:Moti-Example}.

This reveals a fundamental flaw in the goal of existing methods: by trying to recover all missing cross-attention within limited budgets, they saturate the available budget with irrelevant information, crowding out the tokens essential for query accuracy. This evidence casts doubt on whether such an objective is overly ambitious. From a structural standpoint, if a lightweight mechanism could faithfully recover full cross-attention semantics, it would supplant standard Transformer attention, as the combined cost of precomputation and reconstruction would be significantly lower than the full prefill cost \footnote{For a sequence of length $s$ equally partitioned into $N$ equal chunks, the complexity of chunk-wise precomputation is $\mathcal{O}(s^2/N)$, compared to $\mathcal{O}(s^2)$ for full attention.}. However, current methods show no such potential; instead, they behave as degraded versions of the original Transformer, reflecting an inevitable trade-off under limited computational resources.

\subsection{Our Insight: The User Query as a Prophet}
\label{sec:motivation-our-insight}
The above analysis indicates that the missing cross-attention cannot be fully reconstructed in most cases, suggesting that the objective of global attention recovery should be replaced with a more targeted approach. We contend that cross-attention utility is inherently query-contingent: it serves as the bridge between the user's intent and the retrieved evidence. Therefore, recovering attention for query-irrelevant tokens provides marginal utility for the final answer.

While identifying task-relevant intent is typically difficult due to the non-unified representation of queries, RAG systems offer a distinct structural invariant: the query is almost always placed at the prompt’s conclusion. We capitalize on this layout as an opportunity: by treating the terminal query as a 'prophet,' we can extract precise relevance signals to guide selective recomputation. This transforms a structural convention into a powerful mechanism for high-fidelity attention recovery with minimal cost.

\textbf{The predictive power of query-context attention.}  As illustrated in Fig.~\ref{fig:Moti-question-decode}, the subsets attended by the user query exhibit a consistently high overlap ratio with those attended during decoding, a property that remains robust across various models.
This observation suggests that the user query acts as a \emph{prophet}, revealing which parts of the document are critical for the upcoming generation.

\textbf{Advantages}. \ding{182} This property enables targeted recomputation of high-priority tokens guided by the query’s own foresight, rather than blindly reconstructing the entire missing cross-attention, thereby improving generation quality (Sec.~\ref{sec:design-selection}). \ding{183} It also has the potential to directly provide token importance across all layers, avoiding the computational deadlock observed in prior work (Sec.~\ref{sec:design-recomputation}).

\section{Design}
\label{sec:design}
\subsection{Overview of ProphetKV}
\label{sec:design-overview}

\begin{figure}[t]
    \begin{center} 
        \centerline{\includegraphics[width=0.5\columnwidth]{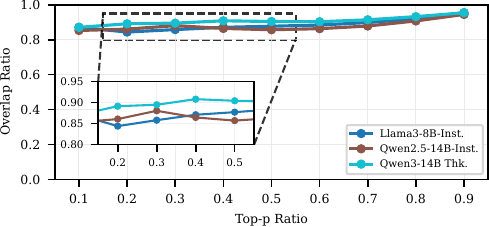}} 
        \vspace{-6pt}
        \caption{The overlap ratio between query-attended tokens and actual critical tokens in the decoding stage. Across various model families, query attention consistently predicts the tokens that will be attended during the decoding stage.}
        \label{fig:Moti-question-decode} 
    \end{center}
    \vspace{-30pt}
\end{figure}

\begin{figure}[t]
    \begin{center}
        \centerline{\includegraphics[width=\columnwidth]{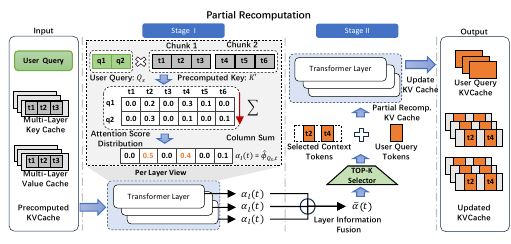}}
        \vspace{-10pt}
        \caption{Overview of the proposed method: utilizing the query as a prophet to guide selective KV cache recomputation.}
        \label{fig:design-overview}
    \end{center}
    \vspace{-30pt}
\end{figure}

Motivated by this insight, we propose ProphetKV, a high-fidelity, position-independent KV cache reuse mechanism based on query-driven selective recomputation. As shown in Fig.~\ref{fig:design-overview}, it employs a dual-stage framework: Stage I generates an evaluation metric based on the user query, and Stage II uses this metric to guide selective recomputation.
Implementing this approach, however, poses two key challenges:

\textbf{Challenge 1: How to quantify token importance via query attention?} 
There needs a quantitative metric for token selection that accurately evaluate the query's attention focus while being computationally efficient.

\textbf{Challenge 2: How to handle attention variability across Transformer layers?}
Attention patterns vary significantly across layers to capture diverse semantic features. This variability makes it difficult to design a selection mechanism that remains robust across the entire depth of LLMs.

\subsection{Query-Driven Token Importance Quantification}
\label{sec:design-selection}
For \textbf{Challenge 1}, we formalize the relationship between query attention and the fidelity of the generated output, bridging the gap between intuition and mathematical rigor.

\textbf{Define the numerical loss function.}
Let $s$ be the number of input tokens, where indices $1 \le t \le s$ and $t>s$ denote input and generated tokens, respectively. Let $Q_s$ be the set of user query tokens. We denote $V_t$ as the $t$-th token's Value tensor and $\Phi_{n,t}$ as the attention weights of the token $n$ to $t$; e.g.,  $\Phi_{Q_s,t}$ means the attention weight of user query tokens to the $t$-th token. For any generated token $n>s$, the output of the attention module can be decomposed as
\begin{equation}
\scriptsize
    \mathrm{AttnOut}_n
    =
    \sum_{t \le n} \Phi_{n, t} V_t
    =
    \sum_{1 \le t \le s} \Phi_{n, t} V_t
    +
    \sum_{s < t \le n} \Phi_{n, t} V_t .
    \label{eq:attention-output-decomposition}
\end{equation}
The second term $\sum_{s < t \le n} \Phi_{n, t} V_t$ corresponds to the contribution of previously generated tokens, and in this work, we focus on the first term, $\sum_{1 \le t \le s} \Phi_{n, t} V_t$, which captures the impact of the input tokens (i.e., reused KV caches).
Guided by insights from Fig.~\ref{fig:Moti-question-decode}, we recognize that the average attention weights of the query tokens, denoted by $\hat{\Phi}_{Q_s,t} = \frac{1}{|Q_s|}\sum_{q \in Q_s} \Phi_{q,t}$, serve as a reliable proxy for $\Phi_{n, t}$. This motivates us to leverage the approximation $\Phi_{n,t} \approx \hat{\Phi}_{Q_s,t}\ (\text{for}\ n > s)$ as a heuristic to derive a quantitative metric for evaluating token importance.

Next, we define the \textit{semantic loss} arising from position-independent KV reuse. In this scenario, K and V tensors are computed for each document in isolation. Such KV caches lack the global context typically provided by cross-document attention, causing both $V_t$ and $\hat{\Phi}_{Q_s,t}$ to become imprecise.
We denote these imprecise counterparts as $V'_t$ ($V_t$ without cross-attention) and $\hat{\Phi}'_{Q_s,t}$ (average attention weights of query tokens to imprecise Key caches), and formulate the resulting semantic loss as follows:
\begin{equation}
\scriptsize
    \mathcal{L}_{\text{semantic}}
    =
    \left\|
    \sum_{1 \le t \le s} \hat{\Phi}_{Q_s,t} V_t
    -
    \sum_{1 \le t \le s} \hat{\Phi}'_{Q_s,t} V'_t
    \right\|_2 .
    \label{eq:semantic-loss}
\end{equation}
Directly optimizing Eq.~\ref{eq:semantic-loss} is computationally intractable because the losses across tokens are coupled. Following prior practices~\cite{delta-attention, scissorhands}, we derive a tractable upper bound using the triangle inequality:
\begin{equation}
\scriptsize
    \mathcal{L}_{\text{semantic}} \le \mathcal{L}
    =
    \sum_{1 \le t \le s}
    \left\|
    \hat{\Phi}'_{Q_s,t} V'_t
    -
    \hat{\Phi}_{Q_s,t} V_t
    \right\|_2 
    \label{eq:loss}
\end{equation}
\textbf{Derive the ideal and practical value functions.}
To mitigate the loss defined in Eq.~\ref{eq:loss} through partial recomputation, we prioritize recomputing KV caches for only some ``critical'' tokens. 
Following prior studies~\cite{cacheblend, kvshare, cacheclip}, we assume that recomputation restores the KV caches of selected tokens to their ground-truth values ($\hat{\Phi}_{Q_s,t}$ and $V_t$) with negligible numerical errors.

Let $\alpha(t)$ be a value function used to identify and rank critical tokens for recomputation. Given a recomputation budget of $p$ ratio, the residual loss incurred by the unrecomputed tokens can be formulated as:
\begin{equation}
\scriptsize
    \mathcal{L}_{\alpha, p} =  \sum_{t \notin \text{TOP}_p(\alpha(t))} \left\|
        \hat{\Phi}'_{Q_s,t} V'_t
        -
        \hat{\Phi}_{Q_s,t} V_t
    \right\|_2,
    \label{eq:target}
\end{equation}
where $\mathrm{TOP}_{p}(\alpha(t))$ denotes the top $p \in [0,1]$ fraction of tokens ranked by $\alpha(t)$.
 
Our aim is to minimize $\mathcal{L}_{\alpha, p}$ through a well-designed $\alpha(t)$.

Obviously, one of the ideal value functions is as follows:
\begin{equation}
\scriptsize
    \alpha_{\text{ideal}}(t)
    =
    \left\|
    \hat{\Phi}_{Q_s,t} V_t
    -
    \hat{\Phi}'_{Q_s,t} V'_t
    \right\|_2 ,
    \label{eq:value-function-ideal}
\end{equation}
because it directly yields the loss for each token to identify ``critical'' tokens, although it cannot be computed in a simple manner.
To derive its tractable formula, we conduct:
\begin{equation}
\scriptsize
    \alpha_{\text{ideal}}(t)
    = \left\|
    (\Delta \hat{\Phi}_{Q_s,t}) V'_t
    +
    (\hat{\Phi}'_{Q_s,t} + \Delta \hat{\Phi}_{Q_s,t}) \Delta V_t
    \right\|_2 ,
    \label{eq:value-function-ideal-derived}
\end{equation}
where $\Delta \hat{\Phi}_{Q_s,t} = \hat{\Phi}_{Q_s,t} - \hat{\Phi}'_{Q_s,t}$ and $\Delta V_t = V_t - V'_t$.

\begin{figure}[t]
    \begin{center}
        \centerline{\includegraphics[width=0.5\columnwidth]{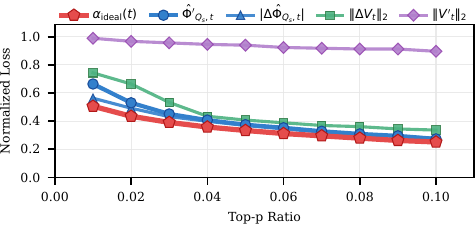}}
        \vspace{-5pt}
        \caption{Comparison of token selection criteria for minimizing semantic loss. $\hat{\Phi}'_{Q_s,t}$ well approximates the ideal value function while remaining observable without a full context prefill.}
        \label{fig:loss-scores-vs-topp}
    \end{center}
    \vspace{-30pt}
\end{figure}

Eq.~\ref{eq:value-function-ideal-derived} identifies four factors influencing recomputation priority: $|\Delta \hat{\Phi}_{Q_s,t}|$, $||V'_t||_2$, $\hat{\Phi}'_{Q_s,t}$, and $||\Delta V_t||_2$. 
To quantify the individual contribution of each one, we conduct an empirical sensitivity analysis by setting $\alpha(t)$ to each candidate in turn and measuring the resulting residual loss $\mathcal{L}_{\alpha, p}$.
The results are shown in Fig.~\ref{fig:loss-scores-vs-topp}, $\alpha(t) = |\Delta \hat \Phi_{Q_s,t}|$ yields the lowest residual loss, followed closely by $\hat{\Phi}'_{Q_s,t}$ and $||\Delta V_t||_2$, while $||V_t'||_2$ performs poorly. 

However, $|\Delta \hat{\Phi}_{Q_s,t}|$ and $||\Delta V_t||_2$ are only obtainable after recomputation, making them unsuitable as recomputation criteria.
Conversely, $\hat{\Phi}'_{Q_s,t}$ is directly observable by a lightweight process that runs Transformers only on the short user query with the imprecise KV caches.
Given this computation convenience, we adopt $\alpha(t) = \hat{\Phi}'_{Q_s,t}$ as our practical proxy for the ideal value function.

\subsection{Overcoming the Deadlock in Token Selection}
\label{sec:design-recomputation} 

Applying selection criteria within the Transformer architecture presents a fundamental challenge (\textbf{Challenge 2}) for prior approaches. 
These approaches~\cite{cacheblend, kvshare} primarily rely on the magnitude of KV tensors to identify critical tokens. However, this strategy is hindered by the layer-wise visibility constraint of Transformers: the KV magnitudes at layer $l$ can only be computed after the full computation of layer $l-1$ is complete.

This creates a structural computational deadlock: to accurately assess token importance at each layer for selective recomputation, these methods would require a full forward pass through all layers. Paradoxically, this amounts to a complete prefill of the entire context—the very computation that KV reuse is intended to avoid—thereby rendering the reuse mechanism ineffective. To circumvent this issue, prior work often assumes layer-wise similarity, namely that tokens deemed important in early layers remain equally important throughout the model. However, this assumption contradicts the fundamental design of Transformers, in which different layers specialize in capturing distinct semantic features at varying levels of abstraction.
Consequently, these methods often struggle to balance selection accuracy with system throughput.

\begin{figure}[t]
    \begin{center} 
        \centerline{\includegraphics[width=0.5\columnwidth]{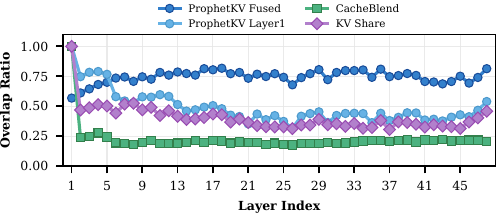}} 
        \vspace{-5pt} 
        \caption{Study on the effectiveness of our fused value function. We evaluate the overlap ratio between the actually selected tokens in each methods and the layer-wise optimal tokens selected by different methods (Appendix~\ref{sec:appendix-selection-strategies}). ProphetKV Layer1, CacheBlend, and KVShare select token only depends on the first layer, whereas ProphetKV Fused considers all layers in token selections.} 
        \label{fig: design-layer-similarity} 
    \end{center}
    \vspace{-30pt}
\end{figure} 

Fortunately, our proposed selection criterion $\hat{\Phi}'_{Q_s,t}$ is inherently immune to this deadlock.
Specifically, $\hat{\Phi}'_{Q_s,t}$ represents the attention weights from the query tokens to the document KV caches across all layers, which provides token importance in all layers without needing a full prefilling.

To ensure robust generalizability (See Appendix~\ref{sec:appendix-layer-difference}), we fuse these per-layer insights into a single global value function. Notably, since attention weights are inherently normalized within each Transformer layer, they provide a consistent scale for comparison across layers. Leveraging this property, we adopt a uniform fusion strategy to avoid the brittleness of manual layer selection as:
\begin{equation}
\scriptsize
    \bar{\alpha}(t) = \frac{1}{L} \sum_{l=1}^{L} \alpha_l(t).
    \label{eq:fused-alpha}
\end{equation}
As shown in Fig.~\ref{fig: design-layer-similarity}, the fused value function achieves much higher overlap with the layer-wise optimal subset, indicating that it reliably captures critical tokens across layers.

Note that our proposed selection criterion $\hat{\Phi}'_{Q_s,t}$ preserves a per-layer computational cost of $\mathcal{O}(|Q_s| \times s)$, which is lower than the $\mathcal{O}(s^2)$ cost required by CacheBlend or KVShare ( Appendix~\ref{sec:appendix-selection-strategies}) in the common case where $|Q_s| \ll s$.

\textbf{Putting It Together: The Dual-Stage Recomputation Pipeline.} 
\textbf{Stage I:} We perform a ``lightweight pass'' across all layers. Instead of full attention, we only compute the attention weights of the current query tokens $Q_s$ to the context tokens. For each layer $l$, we use the column sums of these attention weights to calculate the value function $\hat{\Phi}'_{Q_s,t}$. \textbf{Stage II:} Then, we compute the fused value function $\bar{\alpha}(t)$ as Eq.~\ref{eq:fused-alpha} and select the top-$k$ most critical tokens. We then perform a complete recomputation across all layers, but only for these selected tokens. Since the selection is fixed across all layers at this stage, the model processes tensors with consistent dimensions, which is compute-friendly for GPU throughput and kernel utilization.

\section{Evaluation} 
\label{sec:evaluation} 

\begin{table*}[t]
\centering
\setlength{\tabcolsep}{2pt}

\scriptsize
\caption{Performance comparison of five approaches on RULER (left side of the table) and LongBench (right side of the table) across various models under a 20\% recomputation ratio. Results at other context lengths are provided in Appendix~\ref{subsec:ruler_results}. The complete datasets for all LongBench tasks are presented in Appendix~\ref{subsec:longbench_results}.}
\vspace{-5pt}
\label{tab:accuracy}
\begin{tabular}{l|ccccccccccccccccccc}
\toprule
Methods & CWE & FWE & MK1 & MQ & MV & Single & QA1 & QA2 & VT & \cellcolor{gray!20}Avg. & WQA & TQA & HQA & NQA & MQue & QMSum & PRetr\_en & PRetr\_zh & \cellcolor{gray!20}Avg. \\
\midrule
\rowcolor{gray!10} \textit{Llama-3.1-8B-Inst.} & 96.90 & 93.67 & 100.00 & 98.75 & 100.00 & 100.00 & 77.42 & 48.00 & 95.80 & \cellcolor{gray!20}88.82 & 43.14 & 43.31 & 53.35 & 25.50 & 26.20 & 22.19 & 99.67 & 88.18 & \cellcolor{gray!20}50.19 \\
NaiveReuse & 93.00 & 90.33 & 62.00 & 54.75 & 32.00 & 72.00 & 64.42 & 46.00 & 38.40 & \cellcolor{gray!20}61.36 & 34.88 & 44.31 & 42.92 & \textbf{24.48} & 17.25 & 21.53 & 39.33 & 14.00 & \cellcolor{gray!20}29.84 \\
CacheBlend & 93.60 & 89.33 & 77.00 & 82.25 & 73.50 & \textbf{98.67} & 77.17 & 49.00 & 54.60 & \cellcolor{gray!20}77.27 & 39.10 & \textbf{44.79} & 46.75 & 22.77 & 20.91 & 22.60 & 73.67 & 48.00 & \cellcolor{gray!20}39.82 \\
EPIC & \textbf{96.10} & \textbf{93.00} & 76.00 & 67.00 & 50.50 & 97.00 & 70.75 & \textbf{52.00} & 36.20 & \cellcolor{gray!20}70.32 & 40.39 & 44.65 & 47.46 & 24.03 & 17.88 & 23.01 & 65.33 & 20.50 & \cellcolor{gray!20}35.41 \\
KVShare & 92.90 & 87.33 & 73.00 & 79.50 & 67.25 & 97.00 & 67.42 & 45.00 & 51.40 & \cellcolor{gray!20}73.47 & 36.20 & 44.22 & 45.48 & 23.55 & 20.50 & 22.76 & 65.67 & 51.00 & \cellcolor{gray!20}38.67 \\
\rowcolor{blue!10} ProphetKV & 95.50 & 90.67 & \textbf{99.00} & \textbf{98.50} & \textbf{100.00} & 97.67 & \textbf{77.33} & 51.00 & \textbf{67.00} & \cellcolor{blue!20}\textbf{84.71} & \textbf{43.21} & 44.12 & \textbf{50.69} & 23.68 & \textbf{24.38} & \textbf{23.35} & \textbf{99.00} & \textbf{98.00} & \cellcolor{blue!20}\textbf{50.80} \\
\midrule
\rowcolor{gray!10} \textit{Qwen2.5-14B-Inst.} & 97.70 & 94.00 & 100.00 & 99.75 & 98.50 & 100.00 & 68.92 & 64.00 & 99.40 & \cellcolor{gray!20}90.28 & 54.45 & 40.57 & 62.11 & 26.78 & 33.56 & 22.94 & 99.67 & 98.79 & \cellcolor{gray!20}54.86 \\
NaiveReuse & 94.00 & 98.00 & 69.00 & 55.75 & 29.25 & 96.00 & 51.00 & 43.00 & 35.60 & \cellcolor{gray!20}62.83 & 17.82 & 9.84 & 24.40 & 15.24 & 2.95 & 19.84 & 28.03 & 8.72 & \cellcolor{gray!20}15.86 \\
CacheBlend & 96.40 & 99.00 & 94.00 & 93.00 & 59.25 & \textbf{99.67} & 60.00 & 59.00 & 58.60 & \cellcolor{gray!20}78.11 & 44.39 & 27.42 & 52.61 & 22.25 & 27.02 & 21.48 & 69.11 & 18.92 & \cellcolor{gray!20}35.40 \\
EPIC & \textbf{97.30} & 98.67 & 88.00 & 91.50 & 45.25 & 98.33 & 64.67 & 55.00 & 41.60 & \cellcolor{gray!20}74.04 & 41.70 & 22.94 & 53.36 & 24.48 & 24.00 & 21.98 & 45.72 & 13.76 & \cellcolor{gray!20}30.99 \\
KVShare & 95.80 & 98.67 & 92.00 & 88.50 & 48.75 & \textbf{99.67} & 63.00 & 53.00 & 39.40 & \cellcolor{gray!20}73.35 & 43.85 & 25.80 & 50.07 & 23.93 & 26.76 & 21.76 & 65.06 & 21.30 & \cellcolor{gray!20}34.82 \\
\rowcolor{blue!10} ProphetKV & 96.00 & \textbf{99.33} & \textbf{97.00} & \textbf{97.00} & \textbf{91.25} & \textbf{99.67} & \textbf{70.17} & \textbf{60.00} & \textbf{95.40} & \cellcolor{blue!20}\textbf{88.60} & \textbf{52.32} & \textbf{39.82} & \textbf{58.13} & \textbf{26.68} & \textbf{35.60} & \textbf{22.51} & \textbf{99.67} & \textbf{92.67} & \cellcolor{blue!20}\textbf{53.43} \\
\midrule
\rowcolor{gray!10} \textit{Qwen-3-14B Thk.} & - & - & 100.00 & 100.00 & 98.00 & 100.00 & 78.75 & 74.00 & 100.00 & \cellcolor{gray!20}91.79 & 73.01 & 46.16 & 74.24 & 25.43 & 49.10 & 20.58 & 100.00 & 100.00 & \cellcolor{gray!20}61.06 \\
NaiveReuse & - & - & 50.00 & 49.25 & 27.75 & 75.00 & 54.33 & 41.00 & 15.80 & \cellcolor{gray!20}43.85 & 25.31 & 37.77 & 26.08 & 7.62 & 3.88 & 5.18 & 24.65 & 8.50 & \cellcolor{gray!20}17.37 \\
CacheBlend & - & - & 72.00 & 77.50 & 57.25 & 98.33 & 70.08 & 64.00 & 64.80 & \cellcolor{gray!20}71.99 & 63.91 & 45.30 & 63.85 & 23.82 & 37.81 & \textbf{19.24} & 73.90 & 73.50 & \cellcolor{gray!20}50.17 \\
EPIC & - & - & 68.00 & 75.25 & 51.50 & 99.00 & 71.08 & 67.00 & 43.80 & \cellcolor{gray!20}67.94 & 62.81 & 45.19 & 61.61 & 22.51 & 37.79 & 18.96 & 66.00 & 47.00 & \cellcolor{gray!20}45.23 \\
KVShare & - & - & 68.00 & 77.25 & 56.50 & \textbf{99.33} & 72.75 & 65.00 & 53.00 & \cellcolor{gray!20}70.64 & 61.84 & 45.26 & 64.95 & 22.61 & 33.26 & 18.49 & 71.78 & 73.00 & \cellcolor{gray!20}48.90 \\
\rowcolor{blue!10} ProphetKV & - & - & \textbf{97.00} & \textbf{96.50} & \textbf{95.75} & 98.67 & \textbf{76.42} & \textbf{72.00} & \textbf{100.00} & \cellcolor{blue!20}\textbf{89.89} & \textbf{70.83} & \textbf{46.53} & \textbf{70.77} & \textbf{25.33} & \textbf{44.12} & 19.07 & \textbf{100.00} & \textbf{99.50} & \cellcolor{blue!20}\textbf{59.52} \\
\bottomrule
\end{tabular}
\begin{tablenotes}
\scriptsize
\small
\item Note: For CWE and FWE with Qwen-3-14B Thk., the model repeats the entire input tokens during generation, exceeding the maximum output length; therefore, these results are excluded. 
\end{tablenotes}
\vspace{-16pt}
\end{table*}

We evaluate ProphetKV's ability to optimize the accuracy-efficiency trade-off by addressing the following questions: (i) \textbf{Accuracy}: Can ProphetKV maintain high accuracy under an aggressive recomputation ratio (\S\ref{sec:evaluation-accuracy})? (ii) \textbf{Efficiency}: What are the practical latency gains in real-world long-context settings (\S\ref{sec:evaluation-efficiency})? (iii) \textbf{Robustness}: How does the method generalize across varying configurations (\S\ref{sec:evaluation-ablation})?

\subsection{Environment Setup}
\label{sec:evaluation-environment}
\textbf{Models and Hardware.} 
We evaluate ProphetKV on three representative LLMs: \textbf{Llama-3.1-8B-Instruct}~\cite{llama-3.1-8B-Instruct}, \textbf{Qwen2.5-14B-Instruct}~\cite{qwen2.5-14B-Instruct}, and \textbf{Qwen-3-14B}~\cite{qwen-3-14B}. To accommodate the long-chain reasoning capabilities of Qwen-3-14B, we set its maximum output length to 4K tokens~\cite{cot, plan-budget}; for the other models, we follow the standard limits established in prior work~\cite {pyramidkv}. All experiments are conducted on a heterogeneous cluster equipped with NVIDIA A100, H100, and L20 GPUs.

\textbf{Benchmarks.} We employ two widely used benchmark suites: \textbf{RULER}~\cite{ruler} (8K context length; see Appendix\ref{subsec:ruler_results} for results with extended lengths) for retrieval-intensive stress testing, and \textbf{LongBench}~\cite{longbench} for reasoning and summarization tasks. Both datasets are partitioned into 512-token segments using LangChain~\cite{langchain}.

\textbf{Baselines.} We compare ProphetKV with three state-of-the-art methods: \textbf{CacheBlend}~\cite{cacheblend}, \textbf{KVShare}~\cite{kvshare}, and \textbf{EPIC}~\cite{epic}. Additionally, a \textbf{Naive Reuse} baseline is included to define the lower bound, as it does not perform any recomputation.

\textbf{Implementation.} We implement ProphetKV and all baselines using the HuggingFace Transformers framework~\cite{huggingface-transformers}. This setup isolates algorithmic performance from effects introduced by specific CUDA kernels (e.g., FlashAttention~\cite{flashattention}, FlexAttention~\cite{flexattention}). Such system-level optimizations are orthogonal to our method and can be incorporated for additional gains. We use \emph{Time to First Token} (TTFT) as the primary metric to capture both prefill and recomputation latency.

\subsection{Accuracy Evaluation}
\label{sec:evaluation-accuracy}
\textbf{Performance Overview.} ProphetKV achieves accuracy comparable to full recomputation (Tab.~\ref{tab:accuracy}). Naive Reuse suffers an accuracy degradation due to the loss of cross-chunk information, and existing baselines exhibit unstable performance across tasks. In contrast, ProphetKV identifies tokens influential for generation. This advantage is pronounced in tasks requiring precise localization of relevant spans under contextual interference, such as multi-value (MV) and multi-query (MQ) tasks. Under a constrained recomputation budget of 20\% tokens, ProphetKV achieves an accuracy improvement of 8.8\%-24.9\% on RULER and 18.6\%-50.9\% on LongBench over prior state-of-the-art methods.

\textbf{Cross-Dataset and Cross-Model Robustness.} Across both RULER and LongBench, ProphetKV outperforms all other baselines. On LongBench, selectively recomputing a small subset of query-relevant tokens preserves long-range semantic coherence even beyond 16K tokens. Moreover, the advantage remains consistent across model scales (8B to 14B) and types (instruction-tuned vs. reasoning-oriented), demonstrating that the query-driven mechanism generalizes across diverse datasets and architectural scaling laws.

\begin{figure*}[t]
    \begin{center}
        \centerline{\includegraphics[width=\textwidth]{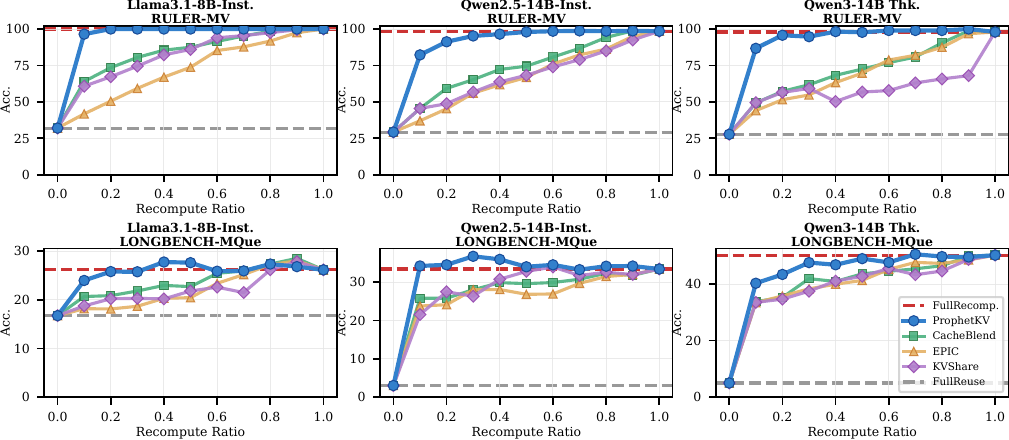}}
        \vspace{-5pt}
        \caption{
            Accuracy results on RULER MultiValue and LongBench Musique across different models and recomputation ratios.
        }
        \label{fig: eval-ratio}
    \end{center}
    \vspace{-30pt}
\end{figure*}

\subsection{Efficiency Evaluation}
\label{sec:evaluation-efficiency}

As shown in Tab.~\ref{tab:ttft}, ProphetKV achieves up to a $5\times$ speedup over full recomputation at a $20\%$ recomputation ratio for both 8K and 16K contexts, where the computation is dominated by attention over long sequences. For 4K contexts, however, such speedups are not observed for all methods, as fixed system overheads (e.g., kernel launch and cache management) become more pronounced and limit the achievable acceleration. Compared to existing baselines, ProphetKV matches EPIC in efficiency and outperforms CacheBlend, thanks to its lightweight first-stage user-query-to-context attention, in contrast to CacheBlend’s heavier first-layer selection.
Consequently, ProphetKV offers a superior trade-off: it matches the speed of the fastest baselines while delivering significantly higher accuracy.

\begin{table}[t]
\centering
\scriptsize
\setlength{\tabcolsep}{2pt}
\caption{TTFT results across different models and context lengths. Each cell shows TTFT in seconds, evaluated on an H100. Complete results, including all baselines, are provided in Appendix~\ref{subsec:full_ttft_results}.}
\vspace{-5pt}
\label{tab:ttft}
\begin{tabular}{l|c|cccc}
\toprule
Model & Context & FullRecomp. & EPIC & ProphetKV & CacheBlend \\
\midrule
\multirow{3}{*}{Llama3-8B-Inst.} & 16K & 5.23 & 1.08 & \textbf{1.13} & 1.29 \\
 & 8K & 1.48 & 0.32 & \textbf{0.35} & 0.38 \\
 & 4K & 0.46 & 0.14 & \textbf{0.15} & 0.14 \\
\midrule
\multirow{3}{*}{Qwen2.5-14B-Inst.} & 16K & 9.94 & 2.03 & \textbf{2.12} & 2.31 \\
 & 8K & 2.70 & 0.58 & \textbf{0.63} & 0.66 \\
 & 4K & 0.88 & 0.23 & \textbf{0.27} & 0.24 \\
\midrule
\multirow{3}{*}{Qwen3-14B Thk.} & 16K & 8.70 & 1.76 & \textbf{1.84} & 2.05 \\
 & 8K & 2.46 & 0.53 & \textbf{0.58} & 0.61 \\
 & 4K & 0.78 & 0.21 & \textbf{0.25} & 0.22 \\
\bottomrule
\end{tabular}
\vspace{-0pt}
\end{table}

\begin{figure}[t]
    \begin{center}
        \centerline{\includegraphics[width=0.5\columnwidth]{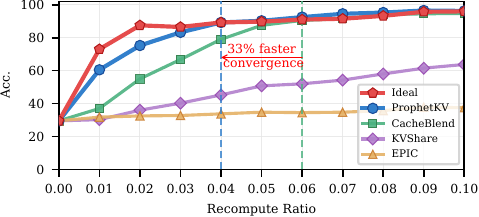}}
        \vspace{-5pt}
        \caption{
            Accuracy of the idealized selection strategy, evaluated on the RULER-MV dataset. The ideal refers to the Eq.\ref{eq:value-function-ideal}.}
        \label{fig: eval-ideal-topp-overlap}
    \end{center}
    \vspace{-26pt}
\end{figure}

\subsection{Ablation Study}
\label{sec:evaluation-ablation}

\textbf{Ablation Study on Selection Strategy.}  
We first investigate the effectiveness of different token selection strategies under an idealized evaluation setting, aiming to isolate performance degradation arising from selection-quality approximations, such as using first-layer information to predict deeper-layer importance. Specifically, we independently execute the full prefill pipeline and the naïve reuse pipeline to collect the exact oracle information required by each method. Using this oracle information, we then apply a unified partial recomputation pipeline---identical to Stage~\uppercase\expandafter{\romannumeral2} of ProphetKV (see Sec.~\ref{sec:design-recomputation})---across all methods. Detailed implementation procedures are provided in Appendix~\ref{appendix:idealized_evaluation}.

As illustrated in Fig.~\ref{fig: eval-ideal-topp-overlap}, the ideal value function (Eq.~\ref{eq:value-function-ideal}) converges rapidly with a recomputation ratio of only 0.02. ProphetKV closely follows this ideal behavior, achieving convergence with a recomputation ratio of 0.04, which is the fastest among all practical methods. In contrast, the best prior method, CacheBlend, requires a recomputation ratio of approximately 0.06 to reach a similar level of overlap. Compared with CacheBlend, ProphetKV achieves approximately a 33\% improvement in convergence speed. These results support our intuition that ProphetKV’s user-query-based selection strategy is more effective than prior approaches, even under an idealized setting simulating unlimited computational resources for token selection.

\begin{figure}[t]
    \begin{center}
        \centerline{\includegraphics[width=0.5\columnwidth]{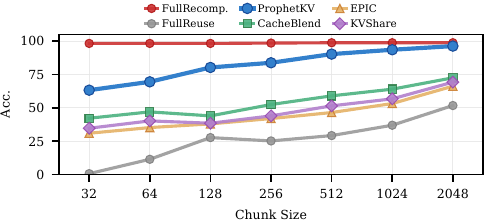}}
        \vspace{-6pt}
        \caption{
            Impact of chunk size on accuracy, evaluated on RULER-MV dataset. The recomputation budget is limited to 20\%.
        }
        \label{fig: eval-chunk-size}
    \end{center}
    \vspace{-33pt}
\end{figure}

\textbf{Ablation Study on Recompute Ratio.} We analyze the robustness of ProphetKV by sweeping the recomputation ratio from 0.0 to 1.0 on RULER MultiValue (retrieval-intensive) and LongBench Musique (reasoning-intensive). As illustrated in Fig.~\ref{fig: eval-ratio}, ProphetKV exhibits a significantly steeper accuracy recovery curve than the baselines. It incurs less than a 5\% accuracy drop with only 10\%--30\% recomputation, whereas baselines typically require 40\%--80\% to achieve similar accuracy. Notably, on RULER-MV, ProphetKV attains near-complete accuracy recovery at a recomputation ratio of 0.2, while baseline methods exhibit substantially slower accuracy recovery as the recomputation ratio increases. These results confirm that ProphetKV effectively prioritizes the most influential tokens, making it highly robust under constrained computational budgets.

\textbf{Ablation Study on Chunk Size.} We further evaluate the impact of chunk size on the accuracy of ProphetKV. As illustrated in Fig.~\ref{fig: eval-chunk-size}, ProphetKV consistently achieves the highest accuracy across all evaluated chunk sizes. In particular, it achieves near-lossless accuracy with chunk sizes above 512 tokens, a common setting in RAG scenarios~\cite{epic}. Accuracy for all methods degrades as chunk size decreases, since smaller chunks lead to more missing cross-chunk attention.

\section{Conclusion}
\label{sec:conclusion}
We proposed ProphetKV, a high-fidelity,  position-independent KV cache reuse mechanism for long-context RAG scenarios.
It leverages query-driven selective recomputation to recover task-critical cross-attention and mitigates the accuracy loss observed in prior work with minimal overhead.
Extensive experiments show that ProphetKV significantly improves accuracy compared to SOTA approaches.

\newpage
\bibliography{example_paper}
\bibliographystyle{plain}

\newpage
\appendix
\onecolumn

\section{Pseudocode of Our Method}
\label{sec:appendix-pseudocode}
Algorithm~\ref{alg:dual_stage_partial_recomputation} presents the dual-stage partial recomputation procedure adopted in ProphetKV. The algorithm takes the precomputed KV caches from multiple chunks across all layers, along with the user query tokens $Q_s$, as input and aims to selectively recover task-critical cross-attention semantics with minimal recomputation cost. In the first stage, the algorithm computes a user-query-aware importance score for each context token at each layer by measuring the attention induced by the user query, resulting in a layer-specific value function $\alpha_l(t)$.

Since attention patterns can vary significantly across layers, directly selecting recomputation tokens independently at each layer leads to unstable and inefficient behavior. To address this issue, the second stage aggregates the layer-wise importance scores into a fused score $\bar{\alpha}(t)$, from which a unified set of top-$p$ tokens is selected for recomputation. Given this unified token set, the KV cache is recomputed independently at each layer, while the remaining cached KV entries are reused directly.

\begin{algorithm}[ht]
    \caption{Dual-Stage Partial Recomputation Algorithm}
    \label{alg:dual_stage_partial_recomputation}
    \begin{algorithmic}
      \STATE {\bfseries Input:} Precomputed KV cache $\{K_{1:C_1}^{\prime(l)}, V_{1:C_1}^{\prime(l)}\}_{l=1}^L, \{K_{1:C_2}^{\prime(l)}, V_{1:C_2}^{\prime(l)}\}_{l=1}^L, \cdots, \{K_{1:C_n}^{\prime(l)}, V_{1:C_n}^{\prime(l)}\}_{l=1}^L\}$(multiple chunks and layers), user query tokens $Q_s$.
  
      \STATE {\itshape // First Stage: User-Query-Aware Token Selection}  
      \FOR{$l=1$ {\bfseries to} $L$}
        \STATE Concatenate all chunks $K_{1:C}^{\prime(l)} = [K_{1:C_1}^{\prime(l)}, K_{1:C_2}^{\prime(l)}, \cdots, K_{1:C_n}^{\prime(l)}]$ and $V_{1:C}^{\prime(l)} = [V_{1:C_1}^{\prime(l)}, V_{1:C_2}^{\prime(l)}, \cdots, V_{1:C_n}^{\prime(l)}]$.
        \STATE Compute value function $\alpha_l(t) = \hat{\Phi}_{Q_s,t}^{(l)} = \textbf{Softmax}(\frac{Q_s \cdot K_{1:C}^{\prime(l)}}{\sqrt{d_k}})$ for user query tokens at layer $l$
      \ENDFOR
  
      \STATE {\itshape // Second Stage: Multi-Layer Attention Fusion}
      \STATE Fuse value function $\bar{\alpha}(t) \leftarrow \sum_{l=1}^L \alpha_l(t)$
      \STATE Select token indices $T_p \leftarrow \text{Top-}p_{t \le C}\, \bar{\alpha}(t)$
  
      \STATE {\itshape // Layer-wise KV Cache Recomputation (Independent across layers)}
      \FOR{$l=1$ {\bfseries to} $L$}
        \STATE Recompute KV cache $\hat{K}_{T_p}^{(l)}, \hat{V}_{T_p}^{(l)}$ at layer $l$
        \STATE Update layer-wise KV cache:
        \STATE \hspace{1em} $\hat{K}_{1:C}^{(l)} \leftarrow K_{[1:C]\setminus T_p}^{\prime(l)} \cup \hat{K}_{T_p}^{(l)}$
        \STATE \hspace{1em} $\hat{V}_{1:C}^{(l)} \leftarrow V_{[1:C]\setminus T_p}^{\prime(l)} \cup \hat{V}_{T_p}^{(l)}$
        \STATE Calculate user query KV Cache: $\hat{K}_{Q_s}^{(l)}, \hat{V}_{Q_s}^{(l)}$
      \ENDFOR
  
      \STATE {\bfseries Output:} Reconstructed KV cache $\{\hat{K}_{1:C}^{(l)}, \hat{V}_{1:C}^{(l)}\}_{l=1}^L\}$, user query KV Cache $\{\hat{K}_{Q_s}^{(l)}, \hat{V}_{Q_s}^{(l)}\}_{l=1}^L\}$
    \end{algorithmic}
  \end{algorithm}

\section{Full Prompt in Section \ref{sec:motivation-failure}}
We provide the full prompt of the example in Sec.~\ref{sec:motivation-failure} in the following, sentences related to the user query are \underline{underlined}.
\label{sec:appendix-example-prompt}
\begin{promptbox}
    \textbf{Chunk 1(32 Tokens):}  A new coffee shop opened near the central park last week. \underline{John's house is in London}. The city library extended its opening hours to 9 PM daily.

    \textbf{Chunk 2(34 Tokens):} The summer music festival attracted over ten thousand attendees this year. \underline{Alice's house is in Paris}. A new batch of public bicycles was put into use in the urban area. 
    
    \textbf{Chunk 3(33 Tokens):} The downtown art gallery is hosting a modern painting exhibition this week. \underline{Alice stays in John's house on Monday}. The local football team won the regional championship last month.
    
    \textbf{User Query:} In which city does Alice stay on Monday?
\end{promptbox}    

\label{sec:appendix-example-system-prompt}
\begin{promptbox}
    \textbf{prefix:} Answer the question based on the given passages. Only give me the answer and do not output any other words.\verb|\n|\verb|\n|The following are given passages.\verb|\n|

    \textbf{query:} \verb|\n|\verb|\n|Answer the question based on the given passages. Only give me the answer and do not output any other words.\verb|\n|\verb|\n|Question: 

    \textbf{suffix:} \verb|\n|Answer:
\end{promptbox}

We merge the above prompt components into the final input prompt as follows:
\begin{promptbox}
    \{prefix\}\verb|\n|\verb|\n|\{Chunk 1\}\verb|\n|\verb|\n|\{Chunk 2\}\verb|\n|\verb|\n|\{Chunk 3\}\verb|\n|\verb|\n|\{query\}\{User Query\}\{suffix\}
\end{promptbox}

Following a typical LLM inference workflow, we pass the merged prompt to the model for answer generation. The selected token subsets from different methods are exported and highlighted in red in Fig.~\ref{fig:Moti-Example}.

\section{Comparison of Selection Strategies of Prior Works}
\label{sec:appendix-selection-strategies}
We compare the selection strategies of prior works, including CacheBlend~\cite{cacheblend}, KVShare~\cite{kvshare}, and EPIC~\cite{epic}. Specifically, we adopt the same notation as in Sec.~\ref{sec:design-selection}.
\begin{table}[htbp]
\caption{Comparison of selection strategies of prior works.}
\centering
\small
\setlength{\tabcolsep}{2pt}
\begin{tabular}{l|c|c|c}
\toprule
Method & Value Function($\alpha(t)$) & Category & Used low-layer approximation  \\
\midrule
EPIC & -(\text{token distance to chunk start}) & Static & No \\
CacheBlend & $||\Delta V_{t}||_2$ & Dynamic & Yes \\
KV Share & ($\sum_{1 \leq n \leq s} \Phi_{n, t}$) $\cdot ||\Delta V_{t}||_1$ & Dynamic & Yes \\
ProphetKV & $\hat{\Phi}'_{Q_s,t}$ & Dynamic & No \\
\bottomrule
\end{tabular}
\label{tab:selection_strategies}
\end{table}

Notably, from Fig.~\ref{fig: eval-ideal-topp-overlap}, the value function of ProphetKV achieves the fastest convergence speed among all the above methods in an idealized setting without any approximation error, which validates the effectiveness of our selection strategy.

\begin{figure*}[h]
    \centering
    \includegraphics[width=0.32\textwidth]{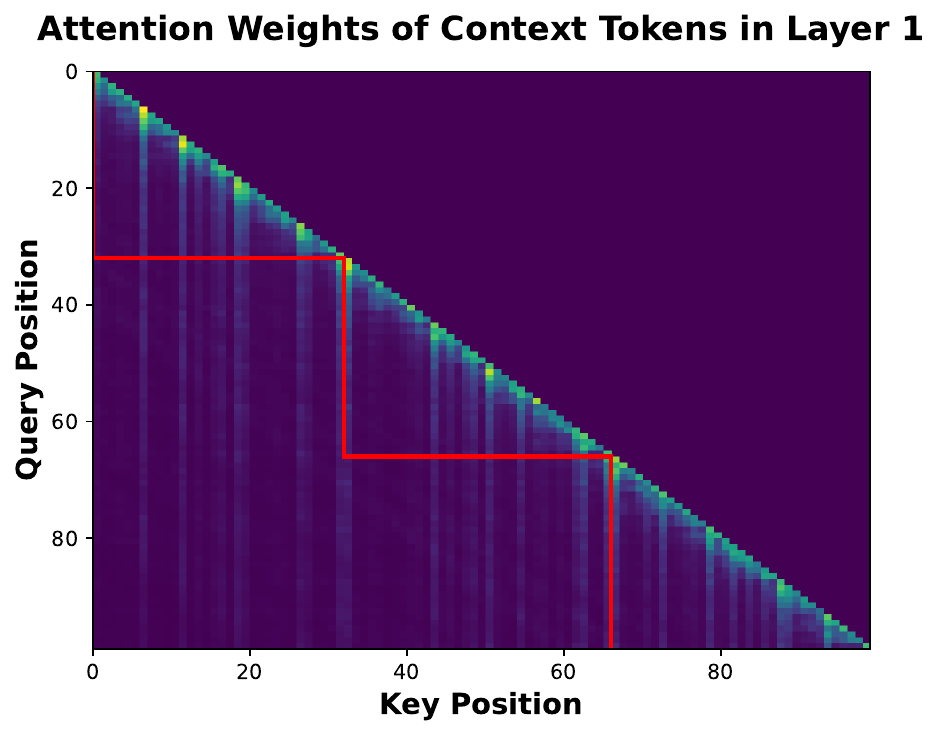}
    \includegraphics[width=0.32\textwidth]{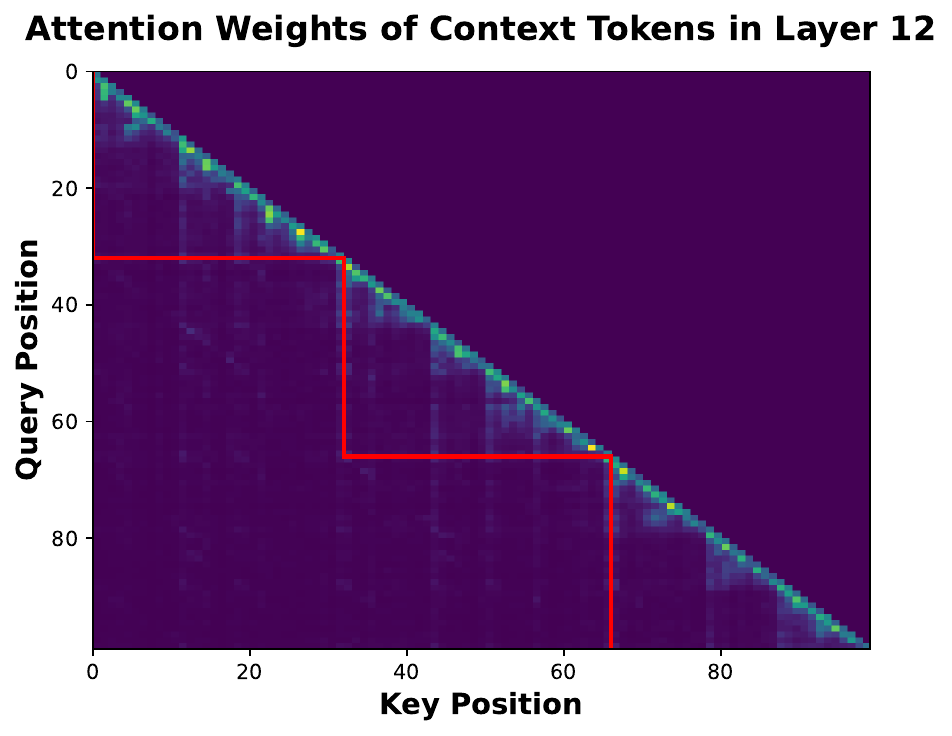}
    \includegraphics[width=0.32\textwidth]{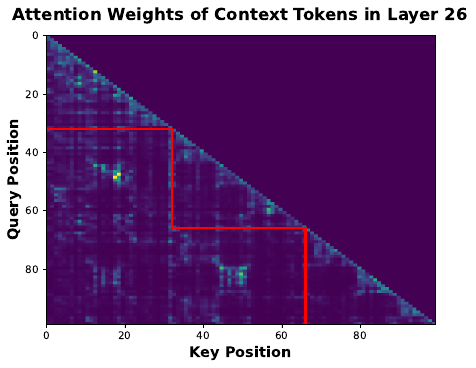}
    \includegraphics[width=0.32\textwidth]{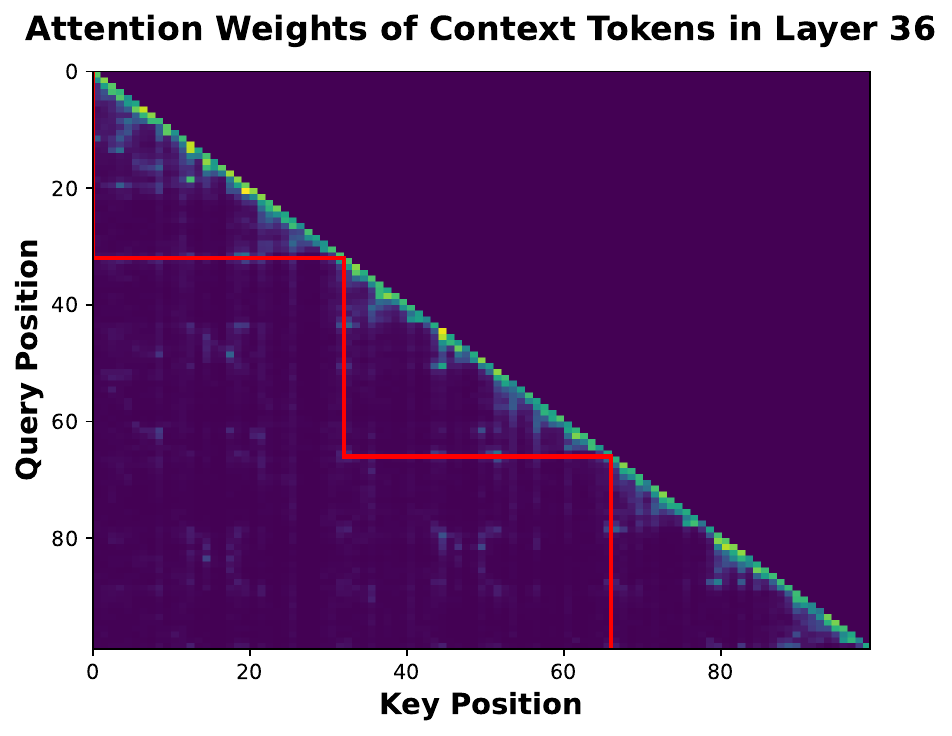}
    \includegraphics[width=0.32\textwidth]{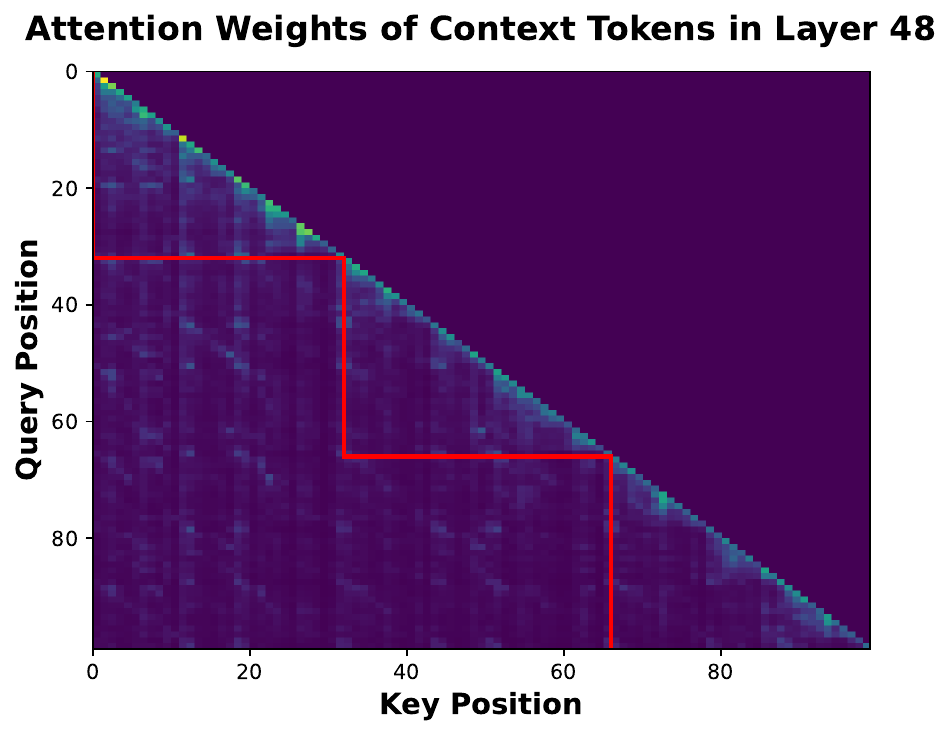}
    \caption{
    Attention weights of context tokens at different layers.
    The cross-attention region is outlined in red in the bottom-left corner.
    }
    \label{fig:appendix-layer-attention}
\end{figure*}

\section{More Details on Inter-Layer Differences}
\label{sec:appendix-layer-difference}

In this section, we use the example prompt in Appendix~\ref{sec:appendix-example-prompt} to illustrate how attention patterns vary across different layers.
We run a full prefill pass for Qwen2.5-14B-Instruct (48 layers in total) and extract the normalized attention weights among context tokens at selected layers to visualize the resulting cross-attention patterns, as shown in Fig.~\ref{fig:appendix-layer-attention}.

As illustrated in Fig.~\ref{fig:appendix-layer-attention}, lower layers (1, 12) focus primarily on nearby tokens, whereas higher layers (26, 36, 48) also attend to distant tokens. And all layers' attention patterns are quite different, leading to the low overlap ratio between the first layer attention weights and the other layer attention weights as shown in Fig.~\ref{fig: design-layer-similarity} in the main text.

\section{Additional Experimental Results}

\subsection{Full TTFT measuring results}\label{subsec:full_ttft_results}

\begin{table}[htbp]
\centering
\scriptsize
\setlength{\tabcolsep}{2pt}
\caption{TTFT results across different models and context lengths. Each cell shows TTFT in seconds.}
\label{tab:full_ttft}
\begin{tabular}{l|c|cccccc}
\toprule
Model & Context & FullRecomp. & EPIC & ProphetKV & CacheBlend & KVShare & NaiveReuse \\
\midrule
\multirow{3}{*}{Llama3-8B-Inst.} & 16K & 5.23 & 1.08 & \textbf{1.13} & 1.29 & 1.32 & 0.09 \\
 & 8K & 1.48 & 0.32 & \textbf{0.35} & 0.38 & 0.38 & 0.05 \\
 & 4K & 0.46 & 0.14 & \textbf{0.15} & 0.14 & 0.14 & 0.04 \\
\midrule
\multirow{3}{*}{Qwen2.5-14B-Inst.} & 16K & 9.94 & 2.03 & \textbf{2.12} & 2.31 & 2.37 & 0.14 \\
 & 8K & 2.70 & 0.58 & \textbf{0.63} & 0.66 & 0.67 & 0.08 \\
 & 4K & 0.88 & 0.23 & \textbf{0.27} & 0.24 & 0.24 & 0.06 \\
\midrule
\multirow{3}{*}{Qwen3-14B Thk.} & 16K & 8.70 & 1.76 & \textbf{1.84} & 2.05 & 2.10 & 0.12 \\
 & 8K & 2.46 & 0.53 & \textbf{0.58} & 0.61 & 0.62 & 0.08 \\
 & 4K & 0.78 & 0.21 & \textbf{0.25} & 0.22 & 0.22 & 0.06 \\
\bottomrule
\end{tabular}
\end{table}

We provide the complete TTFT results across different models and context lengths in Tab.~\ref{tab:full_ttft}. KVShare requires calculating the sum of attention weights as weights for \(\Delta V\), which introduces additional computational overhead compared to CacheBlend. Naive Reuse achieves the lowest TTFT since it does not perform any cross-attention recomputation; however, its accuracy is significantly compromised.

\subsection{Accuracy Results on Ruler with different context lengths}\label{subsec:ruler_results}

\begin{table*}[htbp]
\centering
\setlength{\tabcolsep}{2pt}
\scriptsize
\caption{Performance comparison of different methods on Ruler dataset with 4k context length.}
\label{tab:ruler-4k}
\begin{tabular}{l|cccccccccccc}
\toprule
Methods & CWE & FWE & MK1 & MQ & MV & S1 & S2 & S3 & QA1 & QA2 & VT & \cellcolor{gray!20}Avg. \\
\midrule
\rowcolor{gray!10} \textit{Llama-3.1-8B-Inst.} & 99.60 & 85.67 & 100.00 & 99.25 & 100.00 & 100.00 & 100.00 & 99.00 & 82.08 & 50.00 & 99.60 & \cellcolor{gray!20}92.29 \\
NaiveReuse & 97.80 & 90.67 & 73.00 & 60.25 & 46.05 & 64.00 & 97.98 & 93.00 & 74.42 & 49.00 & 47.80 & \cellcolor{gray!20}72.18 \\
CacheBlend & 97.80 & 95.00 & 88.00 & 88.75 & 81.05 & \textbf{100.00} & \textbf{100.00} & \textbf{98.00} & 78.33 & 50.00 & 55.60 & \cellcolor{gray!20}84.78 \\
EPIC & 99.00 & \textbf{95.67} & 81.00 & 75.75 & 65.79 & 97.00 & 97.98 & 97.00 & 78.08 & 50.00 & 40.80 & \cellcolor{gray!20}79.82 \\
KVShare & 98.90 & 93.67 & 87.00 & 87.50 & 75.26 & 98.00 & \textbf{100.00} & \textbf{98.00} & 72.08 & \textbf{53.00} & 42.00 & \cellcolor{gray!20}82.31 \\
\rowcolor{blue!10} ProphetKV & \textbf{99.30} & 95.33 & \textbf{100.00} & \textbf{99.25} & \textbf{99.74} & \textbf{100.00} & 98.99 & 92.00 & \textbf{82.08} & 51.00 & \textbf{92.60} & \cellcolor{blue!20}\textbf{91.84} \\
\midrule
\rowcolor{gray!10} \textit{Qwen2.5-14B-Inst.} & 99.30 & 92.33 & 100.00 & 100.00 & 99.47 & 100.00 & 100.00 & 100.00 & 75.83 & 64.00 & 100.00 & \cellcolor{gray!20}93.72 \\
NaiveReuse & 98.70 & \textbf{99.00} & 77.00 & 72.25 & 36.84 & 98.00 & \textbf{100.00} & 95.00 & 73.08 & 51.00 & 39.60 & \cellcolor{gray!20}76.41 \\
CacheBlend & \textbf{99.40} & 97.00 & 95.00 & 94.75 & 64.74 & \textbf{100.00} & \textbf{100.00} & \textbf{100.00} & 74.25 & 58.00 & 48.80 & \cellcolor{gray!20}84.72 \\
EPIC & 99.20 & 97.00 & 89.00 & 90.75 & 53.68 & \textbf{100.00} & \textbf{100.00} & \textbf{100.00} & 74.42 & 60.00 & 37.20 & \cellcolor{gray!20}81.93 \\
KVShare & 99.30 & 97.33 & 96.00 & 92.50 & 58.16 & \textbf{100.00} & \textbf{100.00} & \textbf{100.00} & 72.50 & 54.00 & 37.60 & \cellcolor{gray!20}82.49 \\
\rowcolor{blue!10} ProphetKV & 98.40 & 95.67 & \textbf{99.00} & \textbf{97.50} & \textbf{95.00} & \textbf{100.00} & \textbf{100.00} & 99.00 & \textbf{75.58} & \textbf{61.00} & \textbf{95.60} & \cellcolor{blue!20}\textbf{92.43} \\
\midrule
\rowcolor{gray!10} \textit{Qwen-3-14B Thk.} & - & - & 100.00 & 100.00 & 98.68 & 100.00 & 98.99 & 100.00 & 79.75 & 71.00 & 100.00 & \cellcolor{gray!20}94.27 \\
NaiveReuse & - & - & 58.00 & 62.00 & 42.11 & 87.00 & 95.96 & 94.00 & 68.08 & 57.00 & 27.20 & \cellcolor{gray!20}65.71 \\
CacheBlend & - & - & 82.00 & 81.25 & 62.89 & \textbf{100.00} & \textbf{98.99} & \textbf{100.00} & \textbf{82.75} & 64.00 & 65.20 & \cellcolor{gray!20}81.90 \\
EPIC & - & - & 82.00 & 89.25 & 62.63 & \textbf{100.00} & 95.96 & \textbf{100.00} & 78.83 & 65.00 & 50.00 & \cellcolor{gray!20}80.41 \\
KVShare & - & - & 80.00 & 85.50 & 59.47 & \textbf{100.00} & 97.98 & \textbf{100.00} & 78.83 & 61.00 & 55.00 & \cellcolor{gray!20}79.75 \\
\rowcolor{blue!10} ProphetKV & - & - & \textbf{97.00} & \textbf{94.50} & \textbf{88.95} & \textbf{100.00} & \textbf{98.99} & \textbf{100.00} & 78.42 & \textbf{70.00} & \textbf{100.00} & \cellcolor{blue!20}\textbf{91.98} \\
\bottomrule
\end{tabular}
\end{table*}

To further evaluate the robustness of ProphetKV and baseline methods under varying context lengths, we conduct additional accuracy experiments on the RULER dataset with 4K and 16K contexts. These experiments follow the same evaluation protocol as described in the main text, allowing us to systematically assess the impact of context length on retrieval-intensive tasks. As shown in Tab.~\ref{tab:ruler-4k} and Tab.~\ref{tab:ruler-16k}, ProphetKV consistently achieves accuracy comparable to full recomputation across all tasks and models, regardless of context length. Notably, the performance gap between ProphetKV and other baselines remains substantial, particularly as context length increases, underscoring ProphetKV's ability to effectively prioritize and recover critical information in long-context scenarios.

\textbf{Note:} In Tab.~\ref{tab:ruler-16k}, the tokenizer of the Qwen2.5-14B-Instruct model produces longer token sequences for the CWE dataset, resulting in context lengths that exceed the model's maximum limit of 17K tokens. This causes out-of-memory (OOM) errors during evaluation on our 80GB GPU, preventing successful task completion.

\begin{table*}[htbp]
\centering
\setlength{\tabcolsep}{2pt}
\scriptsize
\caption{Performance comparison of different methods on Ruler dataset with 16k context length.}
\label{tab:ruler-16k}
\begin{tabular}{l|cccccccccccc}
\toprule
Methods & CWE & FWE & MK1 & MQ & MV & S1 & S2 & S3 & QA1 & QA2 & VT & \cellcolor{gray!20}Avg. \\
\midrule
\rowcolor{gray!10} \textit{Llama-3.1-8B-Inst.} & 86.00 & 94.00 & 98.00 & 98.50 & 99.75 & 99.00 & 100.00 & 100.00 & 73.83 & 45.00 & 85.60 & \cellcolor{gray!20}89.06 \\
NaiveReuse & 49.60 & \textbf{94.33} & 52.00 & 42.50 & 25.75 & 9.00 & 89.00 & 90.00 & 53.17 & 32.00 & 13.20 & \cellcolor{gray!20}50.05 \\
CacheBlend & 76.20 & 90.33 & 79.00 & 74.75 & 61.25 & 77.00 & 99.00 & 97.00 & 60.17 & 42.00 & 52.00 & \cellcolor{gray!20}73.52 \\
EPIC & \textbf{82.20} & \textbf{94.33} & 69.00 & 64.00 & 36.00 & 25.00 & 98.00 & \textbf{98.00} & 64.50 & 41.00 & 41.60 & \cellcolor{gray!20}64.88 \\
KVShare & 67.80 & 87.33 & 75.00 & 68.50 & 55.25 & 69.00 & \textbf{100.00} & 96.00 & 64.42 & 40.00 & 23.40 & \cellcolor{gray!20}67.88 \\
\rowcolor{blue!10} ProphetKV & 77.00 & 93.33 & \textbf{96.00} & \textbf{94.50} & \textbf{96.25} & \textbf{99.00} & \textbf{100.00} & 94.00 & \textbf{76.50} & \textbf{44.00} & \textbf{54.40} & \cellcolor{blue!20}\textbf{84.09} \\
\midrule
\rowcolor{gray!10} \textit{Qwen2.5-14B-Inst.} & 0.00 & 94.67 & 100.00 & 99.75 & 94.50 & 100.00 & 100.00 & 100.00 & 70.17 & 56.00 & 99.60 & \cellcolor{gray!20}83.15 \\
NaiveReuse & \textbf{0.00} & 97.67 & 57.00 & 42.00 & 26.00 & 97.00 & 97.00 & 83.00 & 29.75 & 36.00 & 30.20 & \cellcolor{gray!20}54.15 \\
CacheBlend & \textbf{0.00} & 97.33 & 88.00 & 89.25 & 43.25 & \textbf{100.00} & 98.00 & 98.00 & 49.42 & 43.00 & 61.40 & \cellcolor{gray!20}69.79 \\
EPIC & \textbf{0.00} & \textbf{98.67} & 78.00 & 79.75 & 34.00 & \textbf{100.00} & 98.00 & 94.00 & 55.42 & 49.00 & 34.80 & \cellcolor{gray!20}65.60 \\
KVShare & \textbf{0.00} & 94.33 & 78.00 & 83.25 & 43.00 & 99.00 & 97.00 & 98.00 & 52.33 & 48.00 & 36.00 & \cellcolor{gray!20}66.26 \\
\rowcolor{blue!10} ProphetKV & \textbf{0.00} & 98.33 & \textbf{98.00} & \textbf{97.00} & \textbf{86.25} & 99.00 & \textbf{99.00} & \textbf{100.00} & \textbf{67.50} & \textbf{57.00} & \textbf{88.80} & \cellcolor{blue!20}\textbf{80.99} \\
\midrule
\rowcolor{gray!10} \textit{Qwen-3-14B Thk.} & - & - & 99.00 & 100.00 & 98.75 & 100.00 & 97.00 & 100.00 & 76.83 & 69.00 & 100.00 & \cellcolor{gray!20}93.40 \\
NaiveReuse & - & - & 25.00 & 10.75 & 12.50 & 11.00 & 43.00 & 33.00 & 12.33 & 22.00 & 4.20 & \cellcolor{gray!20}19.31 \\
CacheBlend & - & - & 60.00 & 63.25 & 45.00 & \textbf{100.00} & 97.00 & 99.00 & 55.75 & 56.00 & 59.20 & \cellcolor{gray!20}70.58 \\
EPIC & - & - & 52.00 & 57.75 & 38.50 & \textbf{100.00} & 92.00 & \textbf{100.00} & 60.08 & 56.00 & 45.00 & \cellcolor{gray!20}66.81 \\
KVShare & - & - & 58.00 & 69.50 & 43.00 & \textbf{100.00} & \textbf{98.00} & 96.00 & 49.42 & 58.00 & 59.40 & \cellcolor{gray!20}70.15 \\
\rowcolor{blue!10} ProphetKV & - & - & \textbf{97.00} & \textbf{97.50} & \textbf{93.50} & \textbf{100.00} & 96.00 & 98.00 & \textbf{78.83} & \textbf{73.00} & \textbf{100.00} & \cellcolor{blue!20}\textbf{92.65} \\
\bottomrule
\end{tabular}
\end{table*}

\subsection{Accuracy Results on other LongBench datasets}\label{subsec:longbench_results}

In the LongBench dataset, certain tasks include extended datasets for more challenging evaluation, such as 2wikimqa and passage\_retrieval\_en. We report results on the extended datasets in Tab.~\ref{tab:accuracy} when available, and present the original results in Tab.~\ref{tab:longbench}. Results for other extended datasets are shown in Tab.~\ref{tab:longbench_extended}.

\begin{table*}[htbp]
\centering
\setlength{\tabcolsep}{2pt}
\scriptsize
\caption{LongBench Results}
\label{tab:longbench}
\begin{tabular}{l|cccccccccccccccc}
\toprule
Methods & WQA & HQA & SAM & DRead & GRep & LCC & MNews & MFQA\_en & MFQA\_zh & PassCnt & PRetr\_en & Qasper & RepoB & TQA & VC & \cellcolor{gray!20}Avg. \\
\midrule
\rowcolor{gray!10} \textit{Llama-3.1-8B-Inst.} & 42.40 & 51.77 & 8.00 & 23.08 & 28.79 & 16.66 & 23.28 & 55.75 & 61.53 & 2.94 & 99.50 & 38.32 & 16.40 & 46.99 & 11.85 & \cellcolor{gray!20}35.15 \\
NaiveReuse & 37.06 & 44.62 & 7.46 & 20.66 & 30.89 & 16.70 & \textbf{24.77} & 41.90 & 46.21 & \textbf{5.93} & 26.50 & 33.64 & 9.74 & \textbf{47.53} & 6.90 & \cellcolor{gray!20}26.70 \\
CacheBlend & 35.24 & 51.23 & 8.19 & \textbf{23.82} & \textbf{31.24} & 17.82 & 24.12 & 49.37 & 52.70 & 5.15 & 60.00 & 37.72 & 13.70 & 46.18 & 6.21 & \cellcolor{gray!20}30.85 \\
EPIC & 37.77 & 46.02 & 7.51 & 22.19 & 31.13 & 17.24 & 24.17 & 48.45 & 51.32 & 4.41 & 50.50 & \textbf{38.86} & 12.34 & 46.47 & 5.10 & \cellcolor{gray!20}29.57 \\
KVShare & 36.65 & 48.63 & 8.13 & 21.27 & 30.83 & \textbf{17.89} & 23.31 & 48.32 & 53.47 & 5.15 & 48.00 & 37.78 & 13.21 & 46.47 & 8.97 & \cellcolor{gray!20}29.87 \\
\rowcolor{blue!10} ProphetKV & \textbf{42.09} & \textbf{52.70} & \textbf{9.48} & 21.31 & 29.06 & 17.74 & 23.09 & \textbf{53.39} & \textbf{55.54} & 2.39 & \textbf{96.00} & 36.49 & \textbf{14.25} & 46.67 & \textbf{16.99} & \cellcolor{blue!20}\textbf{34.48} \\
\midrule
\rowcolor{gray!10} \textit{Qwen2.5-14B-Inst.} & 59.71 & 62.82 & 8.60 & 27.97 & 29.93 & 1.75 & 23.21 & 52.18 & 63.93 & 6.20 & 99.00 & 38.06 & 1.37 & 41.59 & 14.80 & \cellcolor{gray!20}35.41 \\
NaiveReuse & 23.87 & 15.55 & 8.85 & 15.07 & 28.08 & 1.71 & \textbf{24.00} & 36.28 & 30.60 & 2.46 & 17.25 & 32.54 & 4.40 & 13.21 & 14.03 & \cellcolor{gray!20}17.86 \\
CacheBlend & 47.27 & 52.86 & 8.83 & 21.42 & 29.73 & 1.78 & 23.19 & 48.65 & 53.49 & \textbf{5.26} & 60.17 & 38.26 & 3.67 & 28.42 & 14.07 & \cellcolor{gray!20}29.14 \\
EPIC & 44.19 & 52.21 & 8.47 & 20.85 & 29.75 & 1.50 & 23.60 & \textbf{50.70} & 49.30 & 1.64 & 41.75 & 38.43 & 3.09 & 21.29 & 14.57 & \cellcolor{gray!20}26.76 \\
KVShare & 43.15 & 50.72 & \textbf{8.92} & 20.62 & 29.94 & \textbf{2.08} & 23.20 & 48.21 & 54.33 & 2.46 & 57.25 & 37.55 & 3.89 & 25.41 & 14.27 & \cellcolor{gray!20}28.13 \\
\rowcolor{blue!10} ProphetKV & \textbf{54.43} & \textbf{58.63} & 8.85 & \textbf{21.78} & \textbf{30.28} & 1.73 & 23.16 & 48.29 & \textbf{58.08} & 5.15 & \textbf{99.00} & \textbf{38.86} & \textbf{4.70} & \textbf{37.95} & \textbf{15.47} & \cellcolor{blue!20}\textbf{33.76} \\
\midrule
\rowcolor{gray!10} \textit{Qwen-3-14B Thk.} & 75.52 & 67.82 & 8.87 & 17.93 & 29.05 & 7.12 & 22.14 & 50.44 & 65.79 & 33.33 & 99.75 & 42.23 & 9.50 & 47.40 & 14.33 & \cellcolor{gray!20}39.41 \\
NaiveReuse & 32.12 & 22.67 & 7.08 & 5.94 & 17.81 & 1.63 & 21.24 & 30.83 & 31.52 & 0.81 & 13.75 & 32.82 & 1.72 & 36.05 & 1.61 & \cellcolor{gray!20}17.17 \\
CacheBlend & 65.76 & 59.45 & 8.41 & 15.39 & \textbf{29.45} & 4.52 & 22.12 & 42.31 & 58.14 & \textbf{3.25} & 66.75 & 40.83 & 3.47 & 47.42 & 2.53 & \cellcolor{gray!20}31.32 \\
EPIC & 63.80 & 56.84 & \textbf{8.93} & 15.01 & 29.34 & 3.25 & \textbf{22.21} & \textbf{45.17} & \textbf{59.10} & \textbf{3.25} & 57.83 & 40.42 & 3.48 & 47.20 & 2.24 & \cellcolor{gray!20}30.54 \\
KVShare & 68.53 & 60.69 & 8.26 & 15.28 & 29.31 & 3.75 & 22.02 & 43.12 & 57.05 & 2.44 & 57.50 & \textbf{41.85} & 2.99 & 48.18 & 2.60 & \cellcolor{gray!20}30.90 \\
\rowcolor{blue!10} ProphetKV & \textbf{72.22} & \textbf{67.74} & 8.51 & \textbf{15.52} & 29.03 & \textbf{6.45} & 21.94 & 44.54 & 56.27 & \textbf{3.25} & \textbf{99.75} & 39.19 & \textbf{4.00} & \textbf{49.31} & \textbf{4.47} & \cellcolor{blue!20}\textbf{34.81} \\
\bottomrule
\end{tabular}
\end{table*}

For challenging cases such as 2wikimqa (WQA) and hotpotqa (HQA), Naive Reuse exhibits a significant accuracy degradation relative to full recomputation. This observation suggests that these tasks require a more comprehensive understanding of global context and cross-chunk interactions.
In contrast, ProphetKV achieves performance comparable to full recomputation in these settings, demonstrating its effectiveness in preserving critical information necessary for accurate response generation.
Conversely, for simpler cases such as gov\_report (GRep), Naive Reuse attains performance on par with full recomputation, indicating that these tasks primarily rely on local context and are less sensitive to cross-chunk interactions. In such scenarios, all partial recomputation methods, including ProphetKV, perform well, highlighting their ability to maintain accuracy while reducing computational overhead.
Finally, certain cases in Qwen-3-14B Thk. (e.g., passage\_count (PassCnt) and repobench-p (RepoB)) suffer from excessively long thinking generation lengths, causing the model to exceed the maximum length of 4K tokens during answer generation. Consequently, the scores for these cases are substantially lower than those of other datasets.

\begin{table*}[htbp]
\centering
\setlength{\tabcolsep}{2pt}
\scriptsize
\caption{LongBench (extended dataset) Results}
\label{tab:longbench_extended}
\begin{tabular}{l|ccccccccc}
\toprule
Methods & SAM & GRep & LCC & MNews & MFQA\_en & PassCnt & Qasper & RepoB & \cellcolor{gray!20}Avg. \\
\midrule
\rowcolor{gray!10} \textit{Llama-3.1-8B-Inst.} & 7.96 & 28.94 & 16.09 & 23.24 & 55.75 & 2.69 & 39.99 & 16.11 & \cellcolor{gray!20}23.85 \\
NaiveReuse & 7.50 & \textbf{31.67} & 14.34 & 22.81 & 41.90 & 7.07 & 35.72 & 9.02 & \cellcolor{gray!20}21.25 \\
CacheBlend & 7.79 & 30.95 & 13.64 & \textbf{23.97} & 49.37 & \textbf{7.12} & 37.77 & 13.39 & \cellcolor{gray!20}23.00 \\
EPIC & 7.31 & 31.08 & 13.04 & 23.64 & 48.45 & 6.11 & \textbf{38.61} & 12.20 & \cellcolor{gray!20}22.55 \\
KVShare & 7.99 & 30.55 & 14.13 & 20.42 & 48.32 & 6.49 & 37.59 & 11.90 & \cellcolor{gray!20}22.17 \\
\rowcolor{blue!10} ProphetKV & \textbf{9.00} & 29.03 & \textbf{16.18} & 23.23 & \textbf{53.39} & 4.73 & 37.53 & \textbf{14.15} & \cellcolor{blue!20}\textbf{23.41} \\
\midrule
\rowcolor{gray!10} \textit{Qwen2.5-14B-Inst.} & 8.69 & 30.19 & 2.97 & 22.23 & 52.18 & 7.95 & 34.47 & 1.60 & \cellcolor{gray!20}20.04 \\
NaiveReuse & \textbf{8.93} & 28.66 & 2.42 & 21.75 & 36.28 & 6.24 & 28.51 & \textbf{5.41} & \cellcolor{gray!20}17.27 \\
CacheBlend & 8.78 & \textbf{30.37} & \textbf{3.09} & 21.60 & 48.65 & 4.56 & \textbf{35.52} & 3.01 & \cellcolor{gray!20}19.45 \\
EPIC & 8.77 & 30.11 & 2.61 & 21.68 & \textbf{50.70} & 5.92 & 34.53 & 2.54 & \cellcolor{gray!20}\textbf{19.61} \\
KVShare & 8.47 & 30.00 & 2.59 & 21.56 & 48.21 & \textbf{7.60} & 34.36 & 2.50 & \cellcolor{gray!20}19.41 \\
\rowcolor{blue!10} ProphetKV & 8.80 & 30.29 & 2.41 & \textbf{21.80} & 48.29 & 5.16 & 34.47 & 3.99 & \cellcolor{blue!20}19.40 \\
\midrule
\rowcolor{gray!10} \textit{Qwen-3-14B Thk.} & 8.32 & 29.59 & 9.07 & 20.86 & 50.44 & 32.09 & 38.55 & 9.30 & \cellcolor{gray!20}24.78 \\
NaiveReuse & 7.03 & 21.89 & 1.27 & 16.33 & 30.83 & 3.92 & 31.15 & 1.95 & \cellcolor{gray!20}14.30 \\
CacheBlend & 8.44 & 29.49 & 2.55 & 20.18 & 42.31 & \textbf{11.15} & 36.78 & 3.16 & \cellcolor{gray!20}19.26 \\
EPIC & 8.29 & 29.52 & 2.61 & 20.22 & \textbf{45.17} & 8.78 & 36.90 & 2.08 & \cellcolor{gray!20}19.20 \\
KVShare & 8.26 & \textbf{29.59} & 1.98 & 20.17 & 43.12 & 7.43 & \textbf{37.96} & 3.11 & \cellcolor{gray!20}18.95 \\
\rowcolor{blue!10} ProphetKV & \textbf{8.62} & 29.42 & \textbf{6.27} & \textbf{20.24} & 44.54 & 6.08 & 37.02 & \textbf{3.20} & \cellcolor{blue!20}\textbf{19.42} \\
\bottomrule
\end{tabular}
\end{table*}


\section{Idealized evaluation setting}\label{appendix:idealized_evaluation}

First, to compute the ideal selection method, we apply Eq.~\ref{eq:value-function-ideal} to each token to obtain an importance score. This requires a full recomputation pass to obtain the accurate Value cache ($V$) and the attention weights from user query tokens to all context tokens ($\hat{\Phi}_{Q_s,t}$). We apply a mean aggregation over the batch and head dimensions, resulting in an attention weight matrix of shape $[\text{Q\_tokens}, \text{cached\_tokens}]$ and a Value matrix of shape $[\text{cached\_tokens}, \text{hidden\_dim}]$ for each layer. We then perform column-wise aggregation along the query-token dimension to obtain an attention weights vector for each context token. Finally, following the layer-wise structure of the Transformer, we aggregate the Value matrices and attention weight vectors across all layers to produce the final $V$ and $\hat{\Phi}_{Q_s,t}$.

Second, we compute the error terms involving $V'$ and $\hat{\Phi}_{Q_s,t}'$ via a reuse evaluation pass, using the same dimension-reduction procedure as above. We compute $\Delta \hat{\Phi}_{Q_s,t}$ as the absolute deviation between $\hat{\Phi}_{Q_s,t}$ and $\hat{\Phi}_{Q_s,t}'$ along the token dimension. Similarly, $\Delta V_t$ is computed as the $\ell_2$ norm of the difference between $V$ and $V'$ for each token along the hidden dimension. Using these quantities, we compute the importance score for each token according to Eq.~\ref{eq:value-function-ideal} and select the top-$p$ tokens for recomputation.

Third, we recompute the KV cache for the selected tokens. We follow the same procedure as CacheBlend, with the key difference that we do not truncate the query or replace the selected tokens' key–value cache after the first layer. Instead, after token embedding, we directly prune the hidden states, positional embedding matrix, and attention mask to retain only the selected tokens for recomputation. During the forward pass, we pass the selected token indices to the attention function to specify how the key–value cache should be replaced.

Notably, these three steps can be integrated into a single model forward pass without requiring multi-turn model loading, and can be efficiently implemented in frameworks such as PyTorch or TensorFlow. We abstract the layer-wise computation into a reusable function and invoke it three times with different inputs to obtain the variables required for ideal selection and recomputation. Special care must be taken when manipulating positional embeddings to ensure that the existing Key cache is correctly aligned with the newly generated cache, as misalignment can lead to erroneous cache replacement. In particular, positional embeddings should be applied to the old Key cache only once, before cache replacement.

\end{document}